%% file: ms.tex
\documentclass{article}
\usepackage{arxiv}

\input{sections/packages.tex}

\title{Perceptual Image Anomaly Detection}

\author{
  Nina Tuluptceva\textsuperscript{1,3}, Bart Bakker\textsuperscript{2}, Irina Fedulova\textsuperscript{1}, Anton Konushin\textsuperscript{3} \\\
  \textsuperscript{1} Philips Research, Moscow, Russia  \\
  \textsuperscript{2} Philips Research, Eindhoven, Netherlands\\
  \textsuperscript{3} Lomonosov Moscow State University, Moscow, Russia \\
  \texttt{{\{nina.tuluptceva, bart.bakker, irina.fedulova\}@philips.com, anton.konushin@graphics.cs.msu.ru}}
}

\begin{document}

\maketitle 

\input{sections/abstract.tex}

\input{sections/introduction.tex}

\input{sections/related_work.tex}
\input{sections/piad.tex}

\input{sections/experiments.tex}

\input{sections/conclusion.tex}

\bibliography{egbib}{}
\bibliographystyle{unsrt}

\end{document}

\maketitle 

\section{Implementation details}

In all experiments images were rescaled to range $[-1, 1]$.
We used the following schedule for training models: $n_{dis}=2$ rounds of discriminator training per one step of generator training (for experiments with LSUN and CelebA we raised $n_{dis}$ to 3). 
During discriminator optimization, the coefficient $\lambda$ for the Gradient Penalty was set to 10 and we also added the regularizer $\mathbb{E}[D(x)^2]$ to the discriminator training objective with a small weight of 0.001, preventing the discriminator outputs to drift far away from zero (as it was done in~\cite{karras2017progressive}).
Networks were trained using the Adam optimizer with betas of (0.5, 0.99) and learning rate 0.0005 for discriminators or 0.0002 for the generator (and the encoder). 

We trained models for 20,000 iterations for the COIL-100 dataset, 50,000 iterations for MNIST and fMNIST, 100,000 for CIFAR-10, and 200,000 for LSUN and CelebA, with a batch size of 32.

\subsection{Relative-perceptual-L1 Loss} 
Computation of the relative-perceptual-L1 loss between two images requires a deep convolutional network that has been pre-trained on an image classification task. Following~\cite{johnson2016perceptual}, we used a 19-layer VGG~\cite{simonyan2014very} network, pre-trained on the Imagenet challenge~\cite{russakovsky2015imagenet}. We computed features using the \textit{relu\_42} layer in the notation of~\cite{johnson2016perceptual} (the activation of the 2nd convolution of the 4th block). Normalizing coefficients $\mu$ and $\sigma$ (means and standard deviations of filter responses) were calculated over a subset of Imagenet as well. For computing the loss on grayscale images, we simply duplicated the channel to get 3-channel images.

\subsection{Gradient-normalizing Weight Policy} 
We used $n_{weight} = 100$ frequency of saving gradient history and updating $\gamma_G$ and $\gamma_E$. In our implementation, instead of computation L2-norm of gradients, we calculated standard deviation. However, since the mean of the derivatives is always close to zero, the standard deviation influences in the same way as L2-norm.

\subsection{Network Architectures} 
\begin{figure}[h!]
\centering
\begin{tabular}{ccc}
\adjustbox{valign=c}{
%---------------------------------------
\begin{tikzpicture} [>=stealth,thick,auto,node distance=2mm,font=\footnotesize]
\tikzset{layer_st/.style={rectangle, draw, minimum width=1.7cm,}}

\node (under_start) []  at (0, 0) {};
\node (start) [fill=black,inner sep=1pt, below = 5mm of under_start]  {};

\node (act1) [layer_st, fill=blue!30, below = 5mm of start] {Leaky ReLU};
\node (conv1) [layer_st, fill=green!30, below = of act1] {$3 \times 3$ Conv};
\node (act2) [layer_st, fill=blue!30, below = of conv1] {Leaky ReLU};
\node (conv2) [layer_st, fill=green!30, below = of act2] {$3 \times 3$ Conv};

\node (finish) [circle, below = 3mm of conv2, draw, inner sep=1pt] {\large +};
\node (below_finish) [below = 5mm of finish] {};

\draw[->] (under_start) to (start);
\draw[->, thin] (start) to (act1);
\draw[->, thin] (act1) to (conv1);
\draw[->, thin] (conv1) to (act2);
\draw[->, thin] (act2) to (conv2);
\draw[->, thin] (conv2) to (finish);
\draw[->] (finish) to (below_finish);

\node (bound_box) [below right = 1mm and 0.5cm of conv1, inner sep=0mm] {};
\draw[] (start) to[out=0, in=90, distance=1cm](bound_box);
\draw[] (bound_box.north) to (bound_box.south);
\draw[->] (bound_box) to[out=270, in=0, distance=0.9cm](finish);

\node [dashed, draw, inner sep=2mm, fit={(start) (finish) (act1) (conv2) (bound_box)}] {};
\end{tikzpicture}
%---------------------------------------
}
& \adjustbox{valign=c} { 
%---------------------------------------
\begin{tikzpicture} [>=stealth,thick,auto,node distance=2mm,font=\footnotesize]
\tikzset{layer_st/.style={rectangle, draw, minimum width=1.7cm,}}

\node (under_start) []  at (0, 0) {};
\node (start) [fill=black,inner sep=1pt, below = 5mm of under_start]  {};

\node (act1) [layer_st, fill=blue!30, below = 5mm of start] {Leaky ReLU};
\node (up1) [layer_st, fill=red!20, below = of act1] {Upsample};
\node (conv1) [layer_st, fill=green!30, below = of up1] {$3 \times 3$ Conv,};
\node (act2) [layer_st, fill=blue!30, below = of conv1] {Leaky ReLU};
\node (conv2) [layer_st, fill=green!30, below = of act2] {$3 \times 3$ Conv};

\node (up2) [layer_st, fill=red!20, right = of up1] {Upsample};
\node (conv_op) [layer_st, fill=green!10, draw=black!50, text=black!30, right = of conv1, below = of up2] {$1 \times 1$ Conv};

\node (finish) [circle, below = 3mm of conv2, draw, inner sep=1pt] {\large +};
\node (below_finish) [below = 5mm of finish] {};

\draw[->] (under_start) to (start);
\draw[->, thin] (start) to (act1);
\draw[->, thin] (act1) to (up1);
\draw[->, thin] (up1) to (conv1);
\draw[->, thin] (conv1) to (act2);
\draw[->, thin] (act2) to (conv2);
\draw[->, thin] (conv2) to (finish);
\draw[->] (finish) to (below_finish);

\draw[->] (start) to[out=0, in=90, distance=0.7cm](up2);
\draw[] (up2) to (conv_op);
\draw[draw=black!50] (conv_op.north) to (conv_op.south);
\draw[->] (conv_op) to[out=270, in=0, distance=1cm](finish);

\node [dashed, draw, inner sep=2mm, fit={(start) (finish) (act1) (conv2) (up2)}] {};
\end{tikzpicture}
%---------------------------------------
} 
& \adjustbox{valign=c} { 
%---------------------------------------
\begin{tikzpicture} [>=stealth,thick,auto,node distance=2mm,font=\footnotesize]
\tikzset{layer_st/.style={rectangle, draw, minimum width=1.7cm,}}

\node (under_start) []  at (0, 0) {};
\node (start) [fill=black,inner sep=1pt, below = 5mm of under_start]  {};

\node (act1) [layer_st, fill=blue!30, below = 5mm of start] {Leaky ReLU};
\node (conv1) [layer_st, fill=green!30, below = of act1] {$3 \times 3$ Conv};
\node (act2) [layer_st, fill=blue!30, below = of conv1] {Leaky ReLU};
\node (conv2) [layer_st, fill=green!30, below = of act2] {$3 \times 3$ Conv};
\node (pool1) [layer_st, fill=magenta!20, below = of conv2] {Avg. Pool};

\node (pool2) [layer_st, fill=magenta!20, right = of conv1] {Avg. Pool};
\node (conv_op) [layer_st, fill=green!10, draw=black!50, text=black!30, right = of act2, below = of pool2] {$1 \times 1$ Conv};

\node (finish) [circle, below = 3mm of pool1, draw, inner sep=1pt] {\large +};
\node (below_finish) [below = 5mm of finish] {};

\draw[->] (under_start) to (start);
\draw[->, thin] (start) to (act1);
\draw[->, thin] (act1) to (conv1);
\draw[->, thin] (conv1) to (act2);
\draw[->, thin] (act2) to (conv2);
\draw[->, thin] (conv2) to (pool1);
\draw[->, thin] (pool1) to (finish);
\draw[->] (finish) to (below_finish);

\draw[->] (start) to[out=0, in=90, distance=0.7cm](pool2);
\draw[] (pool2) to (conv_op);
\draw[draw=black!50] (conv_op.north) to (conv_op.south);
\draw[->] (conv_op) to[out=270, in=0, distance=1cm](finish);

\node [dashed, draw, inner sep=2mm, fit={(start) (finish) (act1) (conv2) (pool2)}] {};
%---------------------------------------
\end{tikzpicture}
} \\
(a) \textit{Pre-act. Block} & (b) \textit{BlockUp} & (c) \textit{BlockDown} \\ 
\end{tabular}
\caption{(a) Pre-activation residual block without normalization. (b) Pre-activation residual block that performs upsampling feature resolution. We used nearest-neighbor upsampling and optionally added $1 \times 1$ convolution in the skip-connection. (c) Pre-activation residual block that performs downsampling feature resolution.}
\label{table:res_block}
\end{figure}
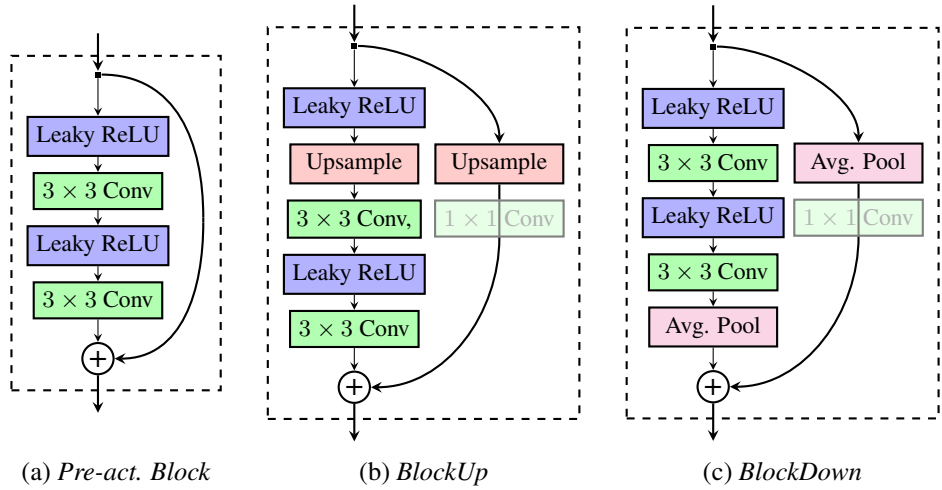

For building the generator $G$, the discriminator $D_X$, and the encoder $E$, we used pre-activation residual blocks~\cite{he2016identity} with two $3\times3$ convolutional layers with the same
number of filters in both convolutions and Leaky ReLU activations with negative slope 0.2. A scheme of the pre-activation residual block is shown in Figure \ref{table:res_block}a. In the generator we used blocks that perform nearest-neighbor upsampling~(\textit{BlockUp}, see Figure~\ref{table:res_block}b). In these blocks, if the convolutions change the number of filters, we additionally put a $1 \times 1$ convolution in the skip-connection path. For the discriminator $D_X$ and the encoder, we used average-pooling downsampling residual blocks~(\textit{BlockDown}, see Figure~\ref{table:res_block}c). We also added \textit{Minibatch Standard Deviation layer}~\cite{karras2017progressive} in the discriminator $D_X$ to improve capturing of variation of image data distribution. 
The discriminator $D_Z$ consists of three fully connected layers with Leaky ReLU activations.

The network architecture for COIL-100, MNIST and fMNIST is shown in Table~\ref{table:model_mnist}, and further, deeper architectures are demonstrated for CIFAR10 in Table~\ref{table:model_cifar10}, for LSUN in Table~\ref{table:model_lsun}, and for CelebA in Table~\ref{table:model_celeba}.

%------------------------------------------------------------------------------------

\begin{table}[h!]
    \centering
	\begin{minipage}[t]{0.5\linewidth}
	\scalebox{0.9}{
		\begin{tabular}[t]{p{2.3cm} p{1cm} p{2cm} p{2cm}}
		\toprule
		\multicolumn{4}{c}{\textbf{Generator $G$}} \\
		\midrule
		& Filters & Kernel size  & Output shape  \\
		\midrule
		Input & & & $32 \times 1 \times 1$ \\
		Convolution & 32 &  $4 \times 4$ (pad 3) & $32 \times 4 \times 4$ \\ 
		BlockUp & 32 & $3 \times 3$ (pad 1) & $32 \times 8 \times 8$ \\
		BlockUp & 32 & $3 \times 3$ (pad 1) & $32 \times 16 \times 16$ \\
		BlockUp & 32 & $3 \times 3$ (pad 1) & $32 \times 32 \times 32$ \\
		Pre-act Block & 32 & $3 \times 3$ (pad 1) & $32 \times 32 \times 32$ \\
		Leaky ReLU & & & $32 \times 32 \times 32$ \\
		Convolution & $\{1, 3\}$ & $1 \times 1$ (pad 0) & $\{1, 3\}  \times 32 \times 32$ \\
		\bottomrule
		\end{tabular}
		}
	\end{minipage}%
	\begin{minipage}[t]{0.5\linewidth}
	\centering
	\scalebox{0.9}{
	    \begin{tabular}[t]{p{2.3cm} p{1cm} p{2cm} p{2cm}}
		\toprule
		\multicolumn{4}{c}{\textbf{Discriminator $D_X$}} \\
		\midrule
		& Filters & Kernel size  & Output shape  \\
		\midrule
		Input & & &  $\{1, 3\} \times 32 \times 32$ \\
		Convolution & 32 & $3 \times 3$ (pad 1) & $32 \times 32 \times 32$ \\
		BlockDown & 32 & $3 \times 3$ (pad 1) & $32 \times 16 \times 16$ \\
		BlockDown & 32 & $3 \times 3$ (pad 1) &$32 \times 8 \times 8$ \\
		BlockDown & 32 & $3 \times 3$ (pad 1) &$32 \times 4 \times 4$ \\
		Minibatch stdev & & & $33 \times 4 \times 4$ \\
		Leaky ReLU & & & $33 \times 4 \times 4$ \\
		Convolution & 32 & $4 \times 4$ (pad 0) & $32 \times 1 \times 1$ \\
		Leaky ReLU & & & $32 \times 1 \times 1$ \\
		Convolution & 1 & $1 \times 1$ (pad 0) & $1 \times 1 \times 1$ \\
		Reshape & & & 1 \\
		\bottomrule
		\end{tabular}
		}
	\end{minipage}
	
	\medskip
	
	\begin{minipage}[t]{0.5\linewidth}
    \centering
	\scalebox{0.9}{
	    \begin{tabular}[t]{p{2.3cm} p{1cm} p{2cm} p{2cm}}
		\toprule
		\multicolumn{4}{c}{\textbf{Encoder $E$}} \\
		\midrule
		& Filters & Kernel size  & Output shape  \\
		\midrule
		Input & & & $\{1, 3\} \times 32 \times 32$ \\
		Convolution & 32 & $3 \times 3$ (pad 1) & $32 \times 32 \times 32$ \\ 
		BlockDown & 32 & $3 \times 3$ (pad 1) & $32 \times 16 \times 16$ \\
		BlockDown & 32 & $3 \times 3$ (pad 1) &$32 \times 8 \times 8$ \\
		BlockDown & 32 & $3 \times 3$ (pad 1) &$32 \times 4 \times 4$ \\
		Leaky ReLU & & & $32 \times 4 \times 4$ \\
		Convolution & 32 & $4 \times 4$ (pad 0) & $32 \times 1 \times 1$ \\
		Leaky ReLU & & & $32 \times 1 \times 1$ \\
		Convolution & 32 & $1 \times 1$ (pad 0) & $32 \times 1 \times 1$ \\
		\bottomrule
		\end{tabular}
		}
	\end{minipage}%
	\begin{minipage}[t]{0.5\linewidth}
    \centering
	\scalebox{0.9}{
	    \begin{tabular}[t]{p{2.3cm} p{3.2cm} p{2cm}}
		\hline
		\multicolumn{3}{c}{\textbf{Discriminator $D_Z$}} \\
		\hline
		& Units & Output shape  \\
		\hline
		Input & & $32 \times 1 \times 1$ \\
		Reshape & & 32 \\
		Dense & 128 & 128 \\
		Leaky ReLU & & 128 \\
		Dense & 128 & 128 \\
		Leaky ReLU & & 128 \\
		Dense & 1 & 1 \\
		\hline
		\end{tabular}
		}
	\end{minipage}	
	
	\medskip
	
	\caption{Network architecture used in MNIST, fMNIST and COIL-100 experiments.}
	\label{table:model_mnist}
% 	\vspace{-5mm}
\end{table}

%-------------------------------------------------------------------------------------
% CIFAR-10

\begin{table}[]
    \centering
	\begin{minipage}[t]{0.5\linewidth}
	\scalebox{0.9}{
		\begin{tabular}[t]{p{2.3cm} p{1cm} p{2cm} p{2cm}}
		\toprule
		\multicolumn{4}{c}{\textbf{Generator $G$}} \\
		\midrule
		 & Filters & Kernel size  & Output shape  \\
		\midrule
		Input & & &  $256 \times 1 \times 1$ \\
		Convolution & 256 &  $4 \times 4$ (pad 3) & $256 \times 4 \times 4$ \\ 
		BlockUp & 256 & $3 \times 3$ (pad 1) & $256 \times 8 \times 8$ \\
		BlockUp & 128 & $3 \times 3$ (pad 1) & $128 \times 16 \times 16$ \\
		BlockUp & 64 & $3 \times 3$ (pad 1) & $64 \times 32 \times 32$ \\
		Pre-act Block & 64 & $3 \times 3$ (pad 1) & $64 \times 32 \times 32$ \\
		Leaky ReLU & & & $64 \times 32 \times 32$ \\
		Convolution & 3 & $1 \times 1$ (pad 0) & $3  \times 32 \times 32$ \\
		\bottomrule
		\end{tabular}
		}
	\end{minipage}%
	\begin{minipage}[t]{0.5\linewidth}
	\centering
	\scalebox{0.9}{
	    \begin{tabular}[t]{p{2.3cm} p{1cm} p{2cm} p{2cm}}
		\toprule
		\multicolumn{4}{c}{\textbf{Discriminator $D_X$}} \\
		\midrule
		& Filters & Kernel size  & Output shape  \\
		\midrule
		Input & & &  $3 \times 32 \times 32$ \\
		Convolution & 32 & $3 \times 3$ (pad 1) & $32 \times 32 \times 32$ \\
		BlockDown & 32 & $3 \times 3$ (pad 1) & $32 \times 16 \times 16$ \\
		BlockDown & 64 & $3 \times 3$ (pad 1) &$64 \times 8 \times 8$ \\
		BlockDown & 128 & $3 \times 3$ (pad 1) &$128 \times 4 \times 4$ \\
		Minibatch stdev & & & $129 \times 4 \times 4$ \\
		Leaky ReLU & & & $129 \times 4 \times 4$ \\
		Convolution & 128 & $4 \times 4$ (pad 0) & $128 \times 1 \times 1$ \\
		Leaky ReLU & & & $128 \times 1 \times 1$ \\
		Convolution & 1 & $1 \times 1$ (pad 0) & $1 \times 1 \times 1$ \\
		Reshape & & & 1 \\
		\bottomrule
		\end{tabular}
		}
	\end{minipage}
	
	\medskip
	
	\begin{minipage}[t]{0.5\linewidth}
    \centering
	\scalebox{0.9}{
	    \begin{tabular}[t]{p{2.3cm} p{1cm} p{2cm} p{2cm}}
		\toprule
		\multicolumn{4}{c}{\textbf{Encoder $E$}} \\
		\midrule
		& Filters & Kernel size  & Output shape  \\
		\midrule
	    Input & & &  $3 \times 32 \times 32$ \\
		Convolution & 64 & $3 \times 3$ (pad 1) & $64 \times 32 \times 32$ \\
		BlockDown & 64 & $3 \times 3$ (pad 1) & $64 \times 16 \times 16$ \\
		BlockDown & 128 & $3 \times 3$ (pad 1) &$128 \times 8 \times 8$ \\
		BlockDown & 256 & $3 \times 3$ (pad 1) &$256 \times 4 \times 4$ \\
		Leaky ReLU & & & $256 \times 4 \times 4$ \\
		Convolution & 256 & $4 \times 4$ (pad 0) & $256 \times 1 \times 1$ \\
		Leaky ReLU & & & $256 \times 1 \times 1$ \\
		Convolution & 256 & $1 \times 1$ (pad 0) & $256 \times 1 \times 1$ \\
		\bottomrule
		\end{tabular}
		}
	\end{minipage}%
	\begin{minipage}[t]{0.5\linewidth}
    \centering
	\scalebox{0.9}{
	    \begin{tabular}[t]{p{2.3cm} p{3.2cm} p{2cm}}
		\hline
		\multicolumn{3}{c}{\textbf{Discriminator $D_Z$}} \\
		\hline
		& Units & Output shape  \\
		\hline
		Input & & $32 \times 1 \times 1$ \\
		Reshape & & 32 \\
		Dense & 1024 & 1024 \\
		Leaky ReLU & & 1024 \\
		Dense & 1024 & 1024 \\
		Leaky ReLU & & 1024 \\
		Dense & 1 & 1 \\
		\hline
		\end{tabular}
		}
	\end{minipage}
	
	\medskip
	
	\caption{Network architecture used in CIFAR10 experiments}
	\label{table:model_cifar10}
\end{table}

%-----------------------------------------------------------------------------------
% LSUN

\begin{table}[]
    \centering
	\begin{minipage}[t]{0.5\linewidth}
	\scalebox{0.9}{
		\begin{tabular}[t]{p{2.3cm} p{1cm} p{2cm} p{2cm}}
		\toprule
		\multicolumn{4}{c}{\textbf{Generator $G$}} \\
		\midrule
		& Filters & Kernel size  & Output shape  \\
		\midrule
		Input & & & $32 \times 4 \times 4$ \\
		Residual Block & 64 & $3 \times 3$ (pad 1) & $64 \times 4 \times 4$ \\
		BlockUp & 64 & $3 \times 3$ (pad 1) & $64 \times 8 \times 8$ \\
		BlockUp & 64 & $3 \times 3$ (pad 1) & $64 \times 16 \times 16$ \\
		BlockUp & 64 & $3 \times 3$ (pad 1) & $64 \times 32 \times 32$ \\
		BlockUp & 64 & $3 \times 3$ (pad 1) & $64 \times 64 \times 64$ \\
		Pre-act Block & 64 & $3 \times 3$ (pad 1) & $64 \times 64 \times 64$ \\
		Leaky ReLU & & & $64 \times 64 \times 64$ \\
		Convolution & 3 & $1 \times 1$ (pad 0) & $3 \times 64 \times 64$ \\
		\bottomrule
		\end{tabular}
		}
	\end{minipage}%
	\begin{minipage}[t]{0.5\linewidth}
	\centering
	\scalebox{0.9}{
	    \begin{tabular}[t]{p{2.3cm} p{1cm} p{2cm} p{2cm}}
		\toprule
		\multicolumn{4}{c}{\textbf{Discriminator $D_X$}} \\
		\midrule
		& Filters & Kernel size  & Output shape  \\
		\midrule
		Input & & &  $3 \times 64 \times 64$ \\
		Convolution & 64 & $3 \times 3$ (pad 1) & $64 \times 64 \times 64$ \\
		BlockDown & 64 & $3 \times 3$ (pad 1) & $64 \times 32 \times 32$ \\
		BlockDown & 64 & $3 \times 3$ (pad 1) & $64 \times 16 \times 16$ \\
		BlockDown & 64 & $3 \times 3$ (pad 1) &$64 \times 8 \times 8$ \\
		BlockDown & 64 & $3 \times 3$ (pad 1) &$64 \times 4 \times 4$ \\
		Minibatch stdev & & & $65 \times 4 \times 4$ \\
		Leaky ReLU & & & $65 \times 4 \times 4$ \\
		Convolution & 64 & $4 \times 4$ (pad 0) & $64 \times 1 \times 1$ \\
		Leaky ReLU & & & $64 \times 1 \times 1$ \\
		Convolution & 1 & $1 \times 1$ (pad 0) & $1 \times 1 \times 1$ \\
		Reshape & & & 1 \\
		\bottomrule
		\end{tabular}
		}
	\end{minipage}
	
	\medskip
	
	\begin{minipage}[t]{0.5\linewidth}
    \centering
	\scalebox{0.9}{
	    \begin{tabular}[t]{p{2.3cm} p{1cm} p{2cm} p{2cm}}
		\toprule
		\multicolumn{4}{c}{\textbf{Encoder $E$}} \\
		\midrule
		& Filters & Kernel size  & Output shape  \\
		\midrule
		Input & & & $3 \times 64 \times 64$ \\
		Convolution & 64 & $3 \times 3$ (pad 1) & $64 \times 64 \times 64$ \\ 
		BlockDown & 64 & $3 \times 3$ (pad 1) & $64 \times 32 \times 32$ \\
		BlockDown & 64 & $3 \times 3$ (pad 1) & $64 \times 16 \times 16$ \\
		BlockDown & 64 & $3 \times 3$ (pad 1) &$64 \times 8 \times 8$ \\
		BlockDown & 64 & $3 \times 3$ (pad 1) &$64 \times 4 \times 4$ \\
		Pre-act Block & & & $64 \times 4 \times 4$ \\
		Leaky ReLU & & & $64 \times 4 \times 4$ \\
		Convolution & 32 & $1 \times 1$ (pad 0) & $32 \times 4 \times 4$ \\
		\bottomrule
		\end{tabular}
		}
	\end{minipage}%
	\begin{minipage}[t]{0.5\linewidth}
    \centering
	\scalebox{0.9}{
	    \begin{tabular}[t]{p{2.3cm} p{3.2cm} p{2cm}}
		\hline
		\multicolumn{3}{c}{\textbf{Discriminator $D_Z$}} \\
		\hline
		& Units & Output shape  \\
		\hline
		Input & & $32 \times 4 \times 4$ \\
		Reshape & &  512 \\
		Dense & 2048 & 2048 \\
		Leaky ReLU & & 2048 \\
		Dense & 2048 & 2048 \\
		Leaky ReLU & & 2048 \\
		Dense & 1 & 1 \\
		\hline
		\end{tabular}
		}
	\end{minipage}	
	
	\medskip
	\vspace{-1mm}
	
	\caption{Network architecture used in LSUN experiments. Residual Block in the generator refers to a residual block that consists of a conv-act-conv sequence.}
	\label{table:model_lsun}
	\vspace{-2mm}
\end{table}

%-----------------------------------------------------------------------------------
% CELEBA

\begin{table}[]
    \centering
	\begin{minipage}[t]{0.5\linewidth}
	\scalebox{0.9}{
		\begin{tabular}[t]{p{2.3cm} p{1cm} p{2cm} p{2cm}}
		\toprule
		\multicolumn{4}{c}{\textbf{Generator $G$}} \\
		\midrule
		& Filters & Kernel size  & Output shape  \\
		\midrule
		Input & & & $64 \times 1 \times 1$ \\
		Convolution & 64 &  $4 \times 4$ (pad 3) & $64 \times 4 \times 4$ \\ 
		BlockUp & 64 & $3 \times 3$ (pad 1) & $64 \times 8 \times 8$ \\
		BlockUp & 64 & $3 \times 3$ (pad 1) & $64 \times 16 \times 16$ \\
		BlockUp & 64 & $3 \times 3$ (pad 1) & $64 \times 32 \times 32$ \\
		BlockUp & 64 & $3 \times 3$ (pad 1) & $64 \times 64 \times 64$ \\
		Pre-act Block & 64 & $3 \times 3$ (pad 1) & $64 \times 64 \times 64$ \\
		Leaky ReLU & & & $64 \times 64 \times 64$ \\
		Convolution & 3 & $1 \times 1$ (pad 0) & $3 \times 64 \times 64$ \\
		\bottomrule
		\end{tabular}
		}
	\end{minipage}%
	\begin{minipage}[t]{0.5\linewidth}
	\centering
	\scalebox{0.9}{
	    \begin{tabular}[t]{p{2.3cm} p{1cm} p{2cm} p{2cm}}
		\toprule
		\multicolumn{4}{c}{\textbf{Discriminator $D_X$}} \\
		\midrule
		& Filters & Kernel size  & Output shape  \\
		\midrule
		Input & & &  $3 \times 64 \times 64$ \\
		Convolution & 64 & $3 \times 3$ (pad 1) & $64 \times 64 \times 64$ \\
		BlockDown & 64 & $3 \times 3$ (pad 1) & $64 \times 32 \times 32$ \\
		BlockDown & 64 & $3 \times 3$ (pad 1) & $64 \times 16 \times 16$ \\
		BlockDown & 64 & $3 \times 3$ (pad 1) &$64 \times 8 \times 8$ \\
		BlockDown & 64 & $3 \times 3$ (pad 1) &$64 \times 4 \times 4$ \\
		Minibatch stdev & & & $65 \times 4 \times 4$ \\
		Leaky ReLU & & & $65 \times 4 \times 4$ \\
		Convolution & 64 & $4 \times 4$ (pad 0) & $64 \times 1 \times 1$ \\
		Leaky ReLU & & & $64 \times 1 \times 1$ \\
		Convolution & 1 & $1 \times 1$ (pad 0) & $1 \times 1 \times 1$ \\
		Reshape & & & 1 \\
		\bottomrule
		\end{tabular}
		}
	\end{minipage}
	
	\medskip
	
	\begin{minipage}[t]{0.5\linewidth}
    \centering
	\scalebox{0.9}{
	    \begin{tabular}[t]{p{2.3cm} p{1cm} p{2cm} p{2cm}}
		\toprule
		\multicolumn{4}{c}{\textbf{Encoder $E$}} \\
		\midrule
		& Filters & Kernel size  & Output shape  \\
		\midrule
		Input & & & $3 \times 64 \times 64$ \\
		Convolution & 64 & $3 \times 3$ (pad 1) & $64 \times 64 \times 64$ \\ 
		BlockDown & 64 & $3 \times 3$ (pad 1) & $64 \times 32 \times 32$ \\
		BlockDown & 64 & $3 \times 3$ (pad 1) & $64 \times 16 \times 16$ \\
		BlockDown & 64 & $3 \times 3$ (pad 1) &$64 \times 8 \times 8$ \\
		BlockDown & 64 & $3 \times 3$ (pad 1) &$64 \times 4 \times 4$ \\
		Leaky ReLU & & & $64 \times 4 \times 4$ \\
		Convolution & 64 & $4 \times 4$ (pad 0) & $64 \times 1 \times 1$ \\
		Leaky ReLU & & & $64 \times 1 \times 1$ \\
		Convolution & 64 & $1 \times 1$ (pad 0) & $64 \times 1 \times 1$ \\
		\bottomrule
		\end{tabular}
		}
	\end{minipage}%
	\begin{minipage}[t]{0.5\linewidth}
    \centering
	\scalebox{0.9}{
	    \begin{tabular}[t]{p{2.3cm} p{3.2cm} p{2cm}}
		\hline
		\multicolumn{3}{c}{\textbf{Discriminator $D_Z$}} \\
		\hline
		& Units & Output shape  \\
		\hline
		Input & & $32 \times 1 \times 1$ \\
		Reshape & & 32 \\
		Dense & 256 & 256 \\
		Leaky ReLU & & 256 \\
		Dense & 256 & 256 \\
		Leaky ReLU & & 256 \\
		Dense & 1 & 1 \\
		\hline
		\end{tabular}
		}
	\end{minipage}	
	
	\medskip
	
	\caption{Network architecture used in CelebA experiments.}
	\label{table:model_celeba}
\end{table}

\bibliography{egbib}{}
\bibliographystyle{unsrt}

%% file: sections/packages.tex
\usepackage{verbatim} % for direct interpretation of latex text 
\usepackage{fancyvrb} % for flexible handling of verbatim text 
\usepackage{xcolor}   % color text
\usepackage{csquotes}
\usepackage{soul}     % for crossed text
\usepackage[]{cite} % for ordering citation
\usepackage{amsmath} % for math formula
\usepackage{amssymb} % for math formula
\usepackage{bm}      % mathbold symbols
\usepackage[makeroom]{cancel} % cross math expression
\usepackage[]{graphicx}
\usepackage{graphbox} % align in images
\graphicspath{{images/}{imgs/}} % path to images
\usepackage{subcaption}
\usepackage{mwe} % to create minimal working examples (MWE) and with dummy images.
\usepackage{tabularx}
\usepackage{multirow}
\usepackage{adjustbox} % align in tables
\usepackage{arydshln}  % dashed separator in tables
\usepackage{makecell}  % to keep spacing of text in tables
\setcellgapes{4pt}     % parameter for the spacing

\usepackage{rotating}
\usepackage{mathtools}
\usepackage{booktabs}
\newcolumntype{L}{>{$}l<{$}}
\newcolumntype{C}{>{$}c<{$}}
\newcolumntype{R}{>{$}r<{$}}

\usepackage{algpseudocode}% http://ctan.org/pkg/algorithmicx
\usepackage{algorithm}% http://ctan.org/pkg/algorithm

\usepackage{tikz}             % diagrams
\usetikzlibrary{positioning}  % relative positioning
\usetikzlibrary{shapes,decorations.pathmorphing} % shapes
\usetikzlibrary{scopes} 
\usetikzlibrary{fit}          % to use nodes inside node
\usetikzlibrary{3d}           % for 3-d vector
\usetikzlibrary{automata, arrows}
\tikzset{>=latex}

%% file: sections/abstract.tex
\begin{abstract}
   We present a novel method for image anomaly detection, where algorithms that use  samples drawn from some distribution of ``normal'' data, aim to detect out-of-distribution (abnormal) samples. 
   Our approach includes a combination of encoder and generator for mapping an image distribution to a predefined latent distribution and vice versa. It leverages Generative Adversarial Networks to learn these data distributions and uses perceptual loss for the detection of image abnormality. To accomplish this goal, we introduce a new similarity metric, which expresses the perceived similarity between images and is robust to changes in image contrast. Secondly, we introduce a novel approach for the selection of weights of a multi-objective loss function (image reconstruction and distribution mapping) in the absence of a validation dataset for hyperparameter tuning.
   After training, our model measures the abnormality of the input image as the perceptual dissimilarity between it and the closest generated image of the modeled data distribution. The proposed approach is extensively evaluated on several publicly available image benchmarks and achieves state-of-the-art performance.
   
\end{abstract}

%% file: sections/introduction.tex
%-------------------------------------------------------------------------
\section{Introduction}
\label{sec:intro}

Anomaly detection is one of the most important problems in a range of real-world settings, including medical applications~\cite{litjens2017survey}, cyber-intrusion detection~\cite{kwon2017survey}, fraud detection~\cite{abdallah2016fraud},  anomaly event detection in videos~\cite{kiran2018overview} and overgeneralization problems of neural networks~\cite{spigler2019denoising}.
Anomaly detection tasks generally involve the use of samples of a  ``normal'' class, drawn from some distribution, to build a classifier that is able to detect ``abnormal'' samples, i.e. \textit{outliers} with respect to the aforementioned distribution.
Although anomaly detection is well-studied in a range of domains, image anomaly detection is still a challenge due to the complexity of distributions over images.

Generative Adversarial Networks~(GANs)~\cite{goodfellow2014generative} present one of the new promising deep anomaly detection approaches.
One network called \textit{the generator} is trained to transform latent vectors, drawn from a latent distribution, to images in such a way that the second network, \textit{the discriminator}, cannot distinguish between real images and generated ones. Thus after training, the generator performs a mapping of the latent distribution to the data distribution. This property has been used~\cite{schlegl2017unsupervised,deecke2018image,perera2019ocgan} to estimate the likelihood of abnormality for an input: if there is a vector in latent space, which after passing through the generator could reconstruct the input object, the object is normal, otherwise it is not. The difference between an input and its closest reconstruction~(\textit{reconstruction error}) is used as an anomaly score for this object.

Although there is a scope of methods that use GAN for anomaly detection, none of them were directly developed for anomaly detection on images.
Usually, they apply the L1-norm or Mean Squared Error (MSE) between the pixels to compute a reconstruction error, which does not correspond to human understanding of the similarity between two images. Another problem of GAN-based approaches is how to find the latent vector that, after passing through the generator, recovers the input object. Previously, 
it was performed by a gradient descent optimization procedure~\cite{schlegl2017unsupervised,deecke2018image}, co-training the generator and \textit{the encoder} that recovers the latent vector~\cite{zenati2018efficient,zenati2018adversarially,perera2019ocgan}.
However, existing techniques are either time-consuming~\cite{schlegl2017unsupervised,deecke2018image} or difficult to train~\cite{zenati2018efficient,zenati2018adversarially}, or consist of complex multi-step learning procedures~\cite{perera2019ocgan}. Another problem is that the complete loss function consists of a sum of many components
with weighting coefficients as hyper-parameters. The lack of a validation set (we do not have any anomaly examples during training), makes it difficult to choose these coefficients.

\begin{figure}[tbp]
    \centering
    \begin{subfigure}[]{0.33\textwidth}
        \centering
        \captionsetup{justification=centering, margin=0cm}
        \begin{subfigure}[]{0.9\textwidth}
            \centering
            \includegraphics[width=1\textwidth]{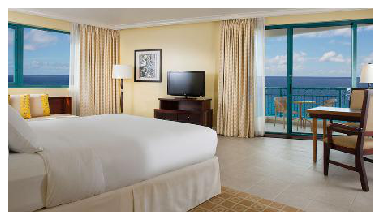}
        \end{subfigure}
        \begin{subfigure}[]{0.9\textwidth}
            \centering
            \includegraphics[width=1\textwidth]{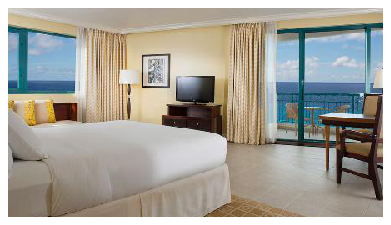}
        \end{subfigure}
    \caption{perc: 4.29 \\ rel-perc-L1: 0.52}
    \label{fig:contrast:a}
    \end{subfigure}%
    \begin{subfigure}[]{0.33\textwidth}
        \centering
        \captionsetup{justification=centering, margin=0cm}
        \begin{subfigure}[]{0.9\textwidth}
            \centering
            \includegraphics[width=1\textwidth]{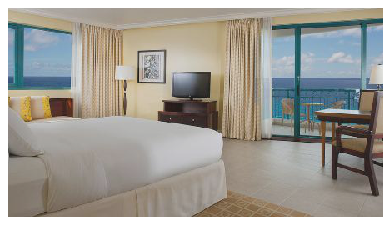}
        \end{subfigure}
        \begin{subfigure}[]{0.9\textwidth}
            \centering
            \includegraphics[width=1\textwidth]{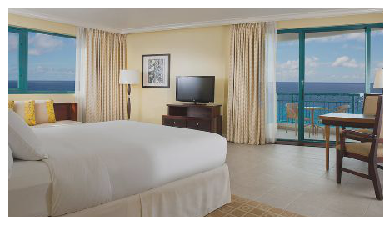}
        \end{subfigure}
    \caption{perc: 3.36 \\ rel-perc-L1: 0.49}
    \label{fig:contrast:b}
    \end{subfigure}%
    \begin{subfigure}[]{0.33\textwidth}
        \centering
        \captionsetup{justification=centering, margin=0cm}
        \begin{subfigure}[]{0.9\textwidth}
            \centering
            \includegraphics[width=1\textwidth]{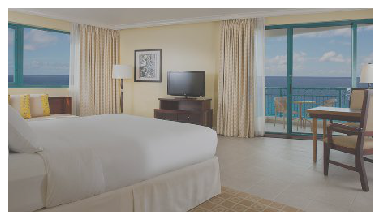}
        \end{subfigure}
        \begin{subfigure}[]{0.9\textwidth}
            \centering
            \includegraphics[width=1\textwidth]{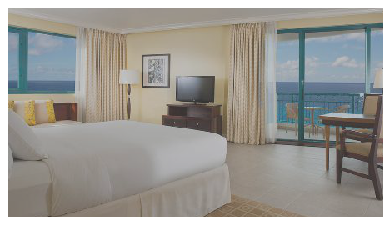}
        \end{subfigure}
    \caption{perc: 2.58 \\ rel-perc-L1: 0.47}
    \label{fig:contrast:c}
    \end{subfigure}
    \caption{Dependence of perceptual loss (\textit{perc}) and proposed relative-perceptual-L1 loss (\textit{rel-perc-L1}) on contrast of images (on an example of LSUN datasets~\cite{xiao2010sun}): (a) Original image and the same one shifted by 5 pixels. (b),~(c)~Images with decreased contrast (by histogram stretching with reduced dynamic range). }
    \label{fig:contrast}

\end{figure}

In our work we propose solutions for each of these three problems:
\begin{enumerate}
\item We developed a new metric that measures the similarity between the perception of two images.
Our metric, called \textit{relative-perceptual-L1 loss}, is based on perceptual loss~\cite{johnson2016perceptual}, but is more robust to noise and changes of contrast of images~(Figure~\ref{fig:contrast}). 
\item We propose a new technique for training an encoder that predicts a latent vector 
jointly with the generator. We construct a loss function in such a way that the encoder predicts a vector \textit{belonging to the latent distribution}, and that the image reconstructed from this vector by the generator \textit{is similar to the input}.
\item We propose a way to choose the weighting coefficients in the complete loss functions for the encoder and the generator.
We base our solution on the norm of the gradients (with respect to network parameters) of each loss function, to balance the contribution of all losses during the training process.
\end{enumerate}

The proposed approach, called Perceptual Image Anomaly Detection (PIAD), allows us to improve performance on several well-known datasets. We experimented with MNIST, Fashion MNIST, COIL-100, CIFAR-10, LSUN and CelebA and made an extensive comparison with a wide range of anomaly detection approaches of  different paradigms.

%% file: sections/related_work.tex
%------------------------------------------------------------------------
\section{Related Work}
\label{sec:related}

Anomaly detection has been extensively studied in a wide range of domains~\cite{chalapathy2019deep}. 
However, anomaly detection on image data is still challenging.  Classical approaches such as explicit modeling of latent space using KDE~\cite{parzen1962estimation}
or One-Class SVM~\cite{chen2001one} which learns a boundary around samples of a normal class, show poor quality when applied to \textit{image} anomaly detection tasks.  Due to the problem of the curse of dimensionality, these algorithms are weak in modeling complex high-dimensional distributions.

Deep autoencoders
play an important role among anomaly detection methods~\cite{sakurada2014anomaly,an2015variational,zhou2017anomaly}. Autoencoders that perform dimension reduction
for normal samples learn some common factors inherent in normal data. Abnormal samples do not contain these factors and thus cannot be accurately reconstructed by autoencoders. 
However, image anomaly detection is still challenging for autoencoders, and usually they are applied only on simple abnormal samples, when the variability of normal images is low.

There are also ``mixed'' approaches that use autoencoders or other deep models for representation learning. 
\textit{GPND}~\cite{pidhorskyi2018generative} leverages an adversarial autoencoder to create a low-dimensional representation and then uses a probabilistic interpretation of the latent space to obtain an anomaly score. The method described in~\cite{abati2019latent} models a latent distribution obtained from a deep autoencoder using an auto-regressive network.
In \textit{Deep SVDD}~\cite{ruff2018deep}, Ruff \textit{et al.} show how to train a one-class classification objective together with deep feature representation.

\begin{figure*}[t]
    \centering
    \begin{subfigure}[t]{0.45\linewidth}
    \centering
    \includegraphics[height=0.13\textheight, trim=0 0 0  0 clip]{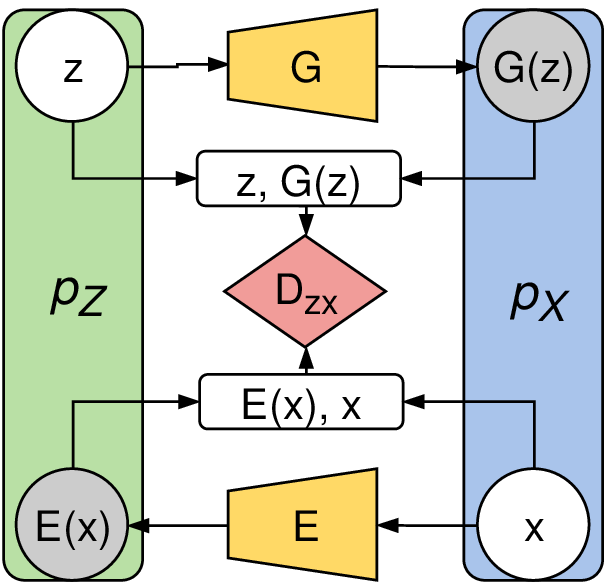}
    \caption{\cite{zenati2018efficient}, based on BIGAN}
    \end{subfigure}%
    \begin{subfigure}[t]{0.45\linewidth}
    \centering
    \includegraphics[height=0.13\textheight, trim=0 0 0  0 clip]{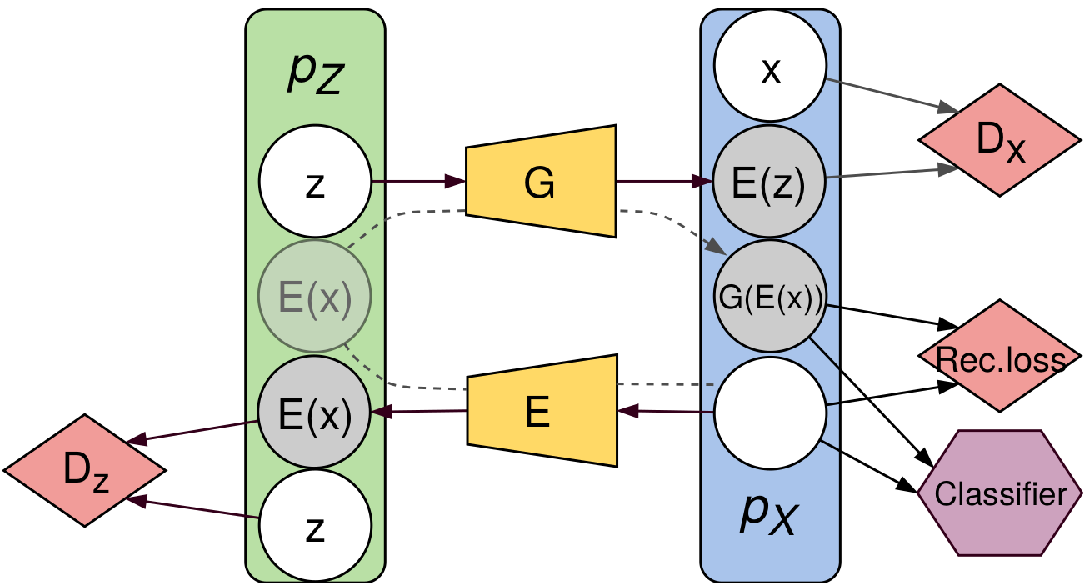}
    \caption{OCGAN~\cite{perera2019ocgan}}
    \end{subfigure}
    
    \begin{subfigure}[t]{0.45\linewidth}
    \centering
    \includegraphics[height=0.13\textheight, trim=0 0 0  0 clip]{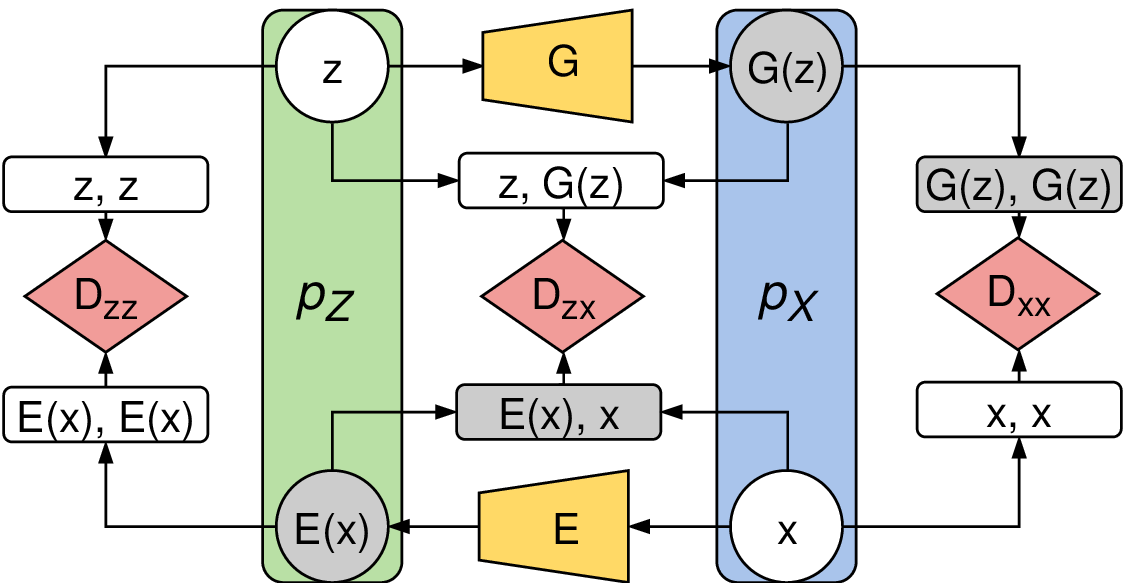}
    \caption{ALAD~\cite{zenati2018adversarially}}
    \end{subfigure}%
    \begin{subfigure}[t]{0.45\linewidth}
    \centering
    \includegraphics[height=0.13\textheight, trim=0 0 0  0 clip]{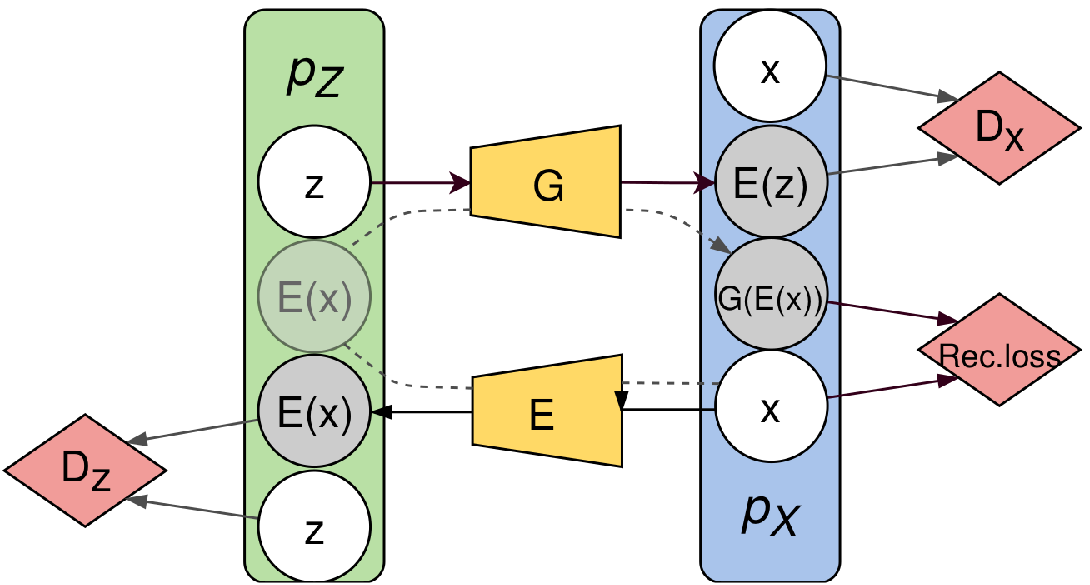}
    \caption{PIAD (ours)}
    \label{fig:app_comp:our}
    \end{subfigure}
    \caption{Comparison of four anomaly detection models. G denotes the generator, E the encoder, $D_*$ the discriminators, ``rec. loss'' reconstruction loss. $p_Z$ denotes the latent distribution, and z (white) its samples. In the same way, $p_X$ is the data distribution and x (white) are data samples.}
    \label{fig:app_comp}
\end{figure*}

GANs~\cite{goodfellow2014generative} created a new branch in the development of image anomaly detection.
GAN-based approaches~\cite{schlegl2017unsupervised,deecke2018image,zenati2018efficient,zenati2018adversarially,perera2019ocgan} differ in two parts:  (i) how to find latent vectors that correspond to the input images, (ii) how to estimate abnormality based on the input image and the reconstructed one. For the second problem, these methods use a linear combination of the L1-norm or the MSE between the input image and the reconstruction~(reconstruction error), and the discriminator's prediction of the reality of the reconstructed image. For the first problem, approaches AnoGAN~\cite{schlegl2017unsupervised} and ADGAN~\cite{deecke2018image} propose to use time-consuming gradient descent optimization of a latent vector.
Other approaches train an encoder to predict a latent vector for each image. Figure~\ref{fig:app_comp} demonstrates the differences between the existing approaches. ALAD~\cite{zenati2018adversarially} and~\cite{zenati2018efficient} train the encoder adversarially: the adversarial loss computed by the discriminator, which takes pairs (image, vector), forces the encoder to predict a latent vector that reconstructs the input image. However, discriminators of such models train with a cross-entropy loss function, which causes an unstable training process. 
The OCGAN model trains a denoising autoencoder. To improve the quality of mapping, authors added 
two discriminators: D$_{X}$
and D$_{Z}$,
and a classifier which searches for hard negative examples (bad generated images). 

%% file: sections/piad.tex
\section{Perceptual Image Anomaly Detection}
\label{sec:approach}

\input{sections/piad/intro.tex}
\input{sections/piad/relative_perceptual_l1.tex}
\input{sections/piad/training_objective.tex}
\input{sections/piad/grad_norm_weight_policy.tex}

%% file: sections/piad/intro.tex
Conceptually the idea of \textit{PIAD} follows the OCGAN. We apply the power of GANs two times, once for building a mapping from the latent space to the image space, and again to create an inverse mapping. A generator and an encoder are trained jointly to satisfy three conditions (see Figure \ref{fig:app_comp:our}): 
\begin{enumerate}
    \item Generator $G$ performs a mapping from latent distribution $p_Z$ to data distribution $p_{X}$; 
    \item Encoder $E$ performs a mapping from $p_X$ to $p_Z$; 
    \item The image which generator $G$ recovers from the latent vector that is predicted by encoder $E$ must be close to the original image (\textit{reconstruction term}): $G(E(x)) \approx x$.
\end{enumerate}

To accomplish conditions 1 and 2 we train the generator and the encoder with adversarial losses. Therefore, two discriminators $D_X$ and $D_Z$
are required. To evaluate the reconstruction term we propose to use our new
\textit{relative-perceptual-L1 loss}.

Ideologically, our approach differs from OCGAN. OCGAN is a denoising autoencoder with highly constrained latent space. On top of reconstruction loss, it uses adversarial loss to ensure that the decoder (the generator in our notation) can reproduce only normal examples. Our approach, however, is based on the power of adversarial loss for mapping two distributions. In practice, OCGAN differs in the classifier component, which helps to find weak places of latent space, which produce not ``normal'' images, but make the whole training process much complicated and multi-steps. Also, we do not add noise to image $x$ before passing it through the encoder. 

In order to train the encoder and generator 
to minimize a multi-objective loss function, 
we propose a new way of setting the weights of the loss functions that equalizes the contribution of each loss
in the training process. Due to the fact that our approach relies on gradients of the parameters of the loss function, we called it \textit{gradient-normalizing weights policy}. 

After training the proposed model on samples of a normal class, we suggest to predict the abnormality of a new example $x$
by evaluating the relative-perceptual-L1 loss between the input $x$ and $G(E(x))$:
\begin{equation}
A(x) = L_{rel-perc-L1}(x, G(E(x))).
\end{equation}

We consider the relative-perceptual-L1 loss in more detail in Section \ref{sec:perc}, the procedure for training models in Section \ref{sec:training} and the gradient-normalizing weights policy in Section \ref{sec:grad_norm}.

%% file: sections/piad/relative_perceptual_l1.tex
\subsection{Relative-perceptual-L1 Loss}
\label{sec:perc}

Features obtained by a neural network, trained on a large dataset for the task of object recognition, can capture high-level image content without binding to exact pixel locations~\cite{gatys2015texture}. In \cite{johnson2016perceptual} Johnson {\it et al.} proposed \textit{content} distance between two images, called \textit{perceptual loss}: this metric computes the MSE between features taken at a deep level of a neural network that has been pre-trained on an object classification task. 

Let $f(x)$ be a feature map obtained from some deep layer of the network on image $x$, and $C \times H \times W$ the shape of this feature map.
Then the \textit{perceptual loss} between image $x$ and $y$ is determined as:
\begin{equation}
L_{perc}(x, y) = \frac{\Vert f(x) - f(y)\Vert_2^2}{C \times H \times W}
\end{equation}

However, perceptual loss is very sensitive to  changes in image contrast. Figure~\ref{fig:contrast} shows three pairs of images: pairs~\ref{fig:contrast:b} and~\ref{fig:contrast:c} have lower contrast than~\ref{fig:contrast:a}.
Perceptual loss drops by 22\% for images~\ref{fig:contrast:b}
compared to~\ref{fig:contrast:a}, although for human supervision the pair~\ref{fig:contrast:b} differs from the pair~\ref{fig:contrast:a} very little. In this way, if we used perceptual loss for computing anomaly score, the model would tend to predict lower contrast images as less abnormal. Another problem is that perceptual loss applies the MSE over features, but the MSE
penalizes the noise in the obtained feature values very heavily. 

\begin{figure}[tbp]
    \vspace{-3mm}
    \centering
    \begin{subfigure}[]{0.44\textwidth}
        \centering
        \includegraphics[height=0.17\textheight]{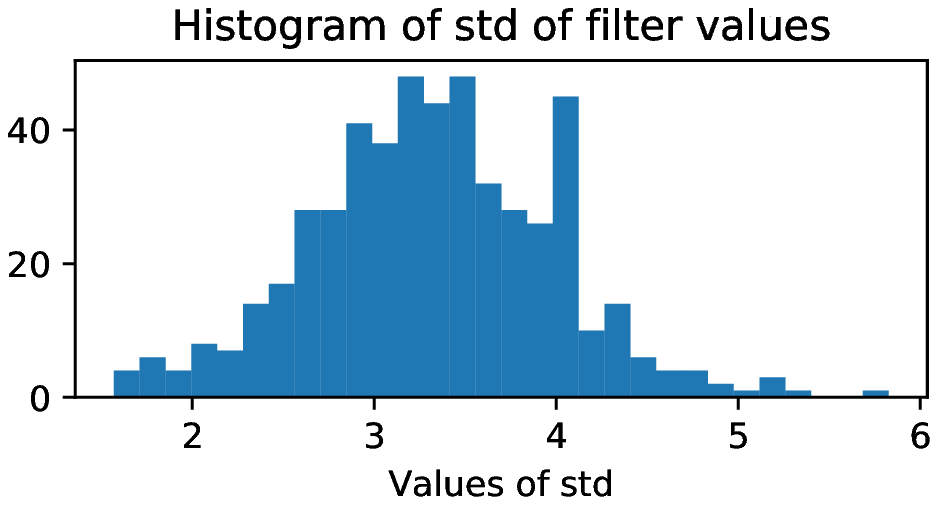}
    \end{subfigure}%
    \begin{subfigure}[]{0.56\textwidth}
         \centering
        \includegraphics[height=0.17\textheight]{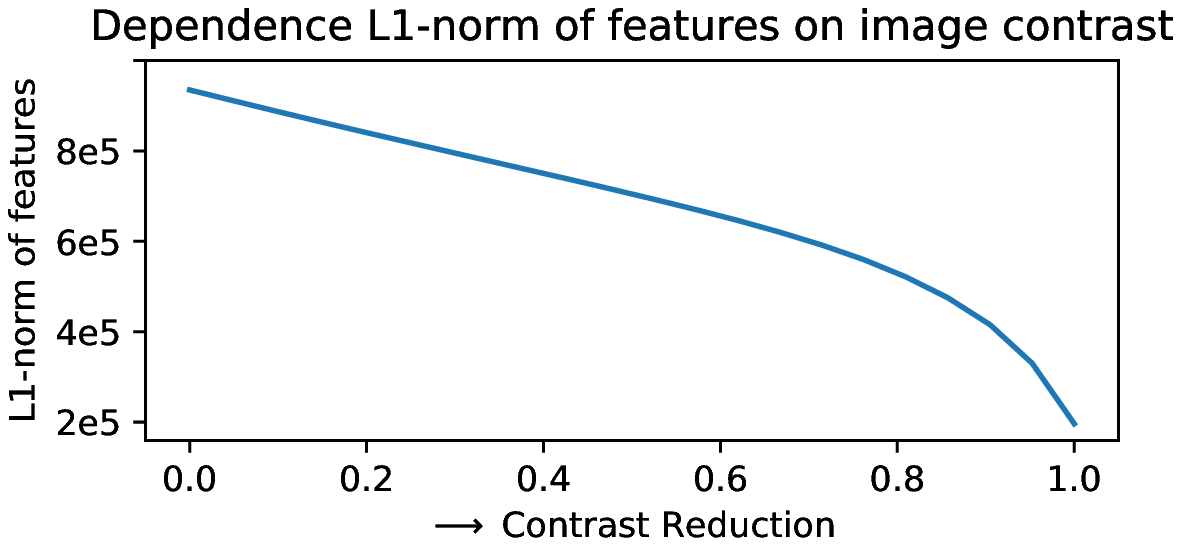}
    \end{subfigure}
    \caption{(left) Histogram of std of responses of 512 filters of a VGG19~\cite{simonyan2014very} layer (the 2nd conv. of the 4-th block), computed over Imagenet~\cite{russakovsky2015imagenet}. (right) Dependence of the L1-norm of the VGG19 features (the 2nd conv. of 4-th block) on image contrast.  Dependence is shown on an image of LSUN dataset (Figure~\ref{fig:contrast}). On the x axis, 0.0 is the original image, 1.0 is completely grey image, between them contrast is linearly decreased (0.2 is corresponds to Figure~\ref{fig:contrast:b}, 0.4 to~\ref{fig:contrast:c}).}
    \label{fig:stats}
\end{figure}

We tackled these problems and propose 
\textit{relative-perceptual-L1} loss, which is robust to contrast and noise.
First of all, we noticed that features obtained at different filters can have a different scatter of values. As an example, Figure~\ref{fig:stats}~(left) shows the standard deviations of filter responses of some deep layer of VGG-19~\cite{simonyan2014very}, computed over Imagenet~\cite{russakovsky2015imagenet}.  We visualize the standard deviations since they indicate the overall value of the features, which are themselves distributed around zero. Standard deviations differ by a factor of 2-3, which means that the contributions per filter vary by a factor 2-3 as well. 
Therefore, as the first step of relative-perceptual-L1, we propose to normalize the obtained deep features by the mean and std of filter responses which are pre-calculated over the large dataset, like Imagenet. 
Secondly, we propose to use the L1-norm instead of the MSE, since the L1-norm is more robust to noise. Thirdly, to make the loss more resistant to contrast, we research how feature values behave under changes of contrast. Figure~\ref{fig:stats}~(right) illustrates this behavior: during the reduction of image contrast, the feature value average decreases. In this way, the absolute error (the difference between features), which is used in perceptual loss, decreases as well for each pair of lower contrast images. Therefore, we propose not to use absolute error, but \textit{relative error}, which measures the ratio of the absolute error of features to the average values of these features. 

Consequently, relative-perceptual-L1 is calculated as follows:
\begin{align}
    &\tilde{f(x)} = (f(x) - \mu) / \sigma, \\
    &L_{rel-perc-L1}(x, y) = \frac{\Vert \tilde{f(x)} -  \tilde{f(y)}\Vert_1}{C \times H \times W} \big/ \frac{\Vert \tilde{f(x)}\Vert_1 }{C \times H \times W} = \frac{\Vert \tilde{f(x)} -  \tilde{f(y)}\Vert_1}{\Vert \tilde{f(x)}\Vert_1, }
\end{align}
where $\mu$, $\sigma$ are the pre-calculated mean and std of filter responses.

%% file: sections/piad/training_objective.tex
\subsection{Training Objective}
\label{sec:training}

To train both discriminators we used the Wasserstein GAN with a Gradient Penalty objective~(WGAN-GP)~\cite{gulrajani2017improved}.
The training of a Wasserstein GAN is more stable than a classical GAN~\cite{goodfellow2014generative} (which was used in~\cite{zenati2018efficient,zenati2018adversarially,perera2019ocgan}), it prevents mode collapse, and does not require a careful searching schedule of generator/discriminator training. Thus, the discriminator $D_X$ learns by
minimizing the following loss:
\begin{align}
\begin{split}
L_{disc}(D_X) &= \mathbb{E}_{z \sim p_Z}[D_X(G(z))] - \mathbb{E}_{x \sim p_X}[D_X(x)] + \lambda \cdot GP(D_X),
\end{split}
\label{eq:dis}
\end{align}
where GP is Gradient Penalty Regularization~\cite{gulrajani2017improved} and $\lambda$ is a weighting parameter. In the same way, $D_Z$ minimizes  $L_{disc}(D_Z)$. Adversarial loss of the generator is
\begin{equation}
L_{adv}(G) =  \mathbb{E}_{x \sim p_X}[D_X(x)]- \mathbb{E}_{z \sim p_Z}[D_X(G(z))].
\end{equation}
Adversarial loss of the encoder $L_{adv}(E)$ is computed in the same way.
Reconstruction loss is measured using the proposed \textit{relative-perceptual-L1} loss:
\begin{equation}
L_{rec}(G, E) = \mathbb{E}_{x \sim p_X}[L_{rel-perc-L1}(x, G(E(x)))].
\end{equation}

Thus, the total objectives for the encoder and generator are as follows:
\begin{align}
L_{total}(G) &= L_{adv}(G) + \gamma_G L_{rec}(G, E), \\
L_{total}(E) &= L_{adv}(E) + \gamma_E L_{rec}(G, E),
\label{eq:total}
\end{align}
where $\gamma_G$ and $\gamma_E$ are weighting parameters.
The training process consists of alternating $n_{dis}$ steps of optimization of the discriminators and one step of optimization of the generator together with the encoder. Parameters $\gamma_G$ and $\gamma_E$ change every $n_{param}$ iterations following our \textit{gradient-normalizing weights policy}. The full training procedure is summarized in Algorithm~\ref{algo:PIAD}. (Steps \textit{``update gradient history''} and \textit{``select weighting parameters''} are explained in detail in the next Section).

\begin{algorithm}[t!]
  \begin{algorithmic}[1]
    \Require N, the total number of iterations. $n_{dis}$, the number of iterations of training the discriminators. $n_{weight}$, frequency of $\gamma_G$ and $\gamma_E$ change. 
    
    \For{$iter = 0, ..., N$}
        \For{$t = 0, ..., n_{dis}$}
            \State Sample $\{x\} \sim p_X$, $\{z\}\sim p_Z$.
            \State Compute $L_{disc}(D_X)$, $L_{disc}(D_Z)$.
            \State Reset gradients; backpropagate $L_{disc}(D_X)$, $L_{disc}(D_Z)$; update $D_X$, $D_Z$.
        \EndFor
        
        \State Sample $\{x\}\sim p_X$, $\{z\} \sim p_Z$ .
        \State Compute $L_{adv}(G)$, $L_{adv}(E)$, $L_{conc}(G, E)$.
        \If{iter \% $n_{weight}$ == 0}
            \For{$loss \in \{L_{adv}(G), L_{adv}(E), L_{rec}(G, E)\}$}
                \State Reset gradients; backpropagate $loss$; update gradient history.
            \EndFor
            \State Select weighting parameters $\gamma_G$ and $\gamma_E$.
        \EndIf
        \State Reset gradients.
        \State Backpropagate $L_{rec}(G, E)$; multiply gradients of $G$  by $\gamma_G$, $E$ by $\gamma_E$.
        \State Backpropagate $L_{adv}(G)$, $L_{adv}(E)$.
        \State Update $G$, $E$.
    \EndFor
    \caption{Training procedure of PIAD. \textit{``Reset gradients''} zeros all stored gradients of networks parameters. \textit{``Backpropagate loss''} computes gradients of loss wrt network parameters and sums them to current stored gradients. \textit{``Update net''} performs one step of gradient descent using stored gradients.}
    \label{algo:PIAD}
\end{algorithmic}
\end{algorithm}

%% file: sections/piad/grad_norm_weight_policy.tex
\subsection{Gradient-normalizing Weight Policy}
\label{sec:grad_norm}

Our objective function 
consists of the sum of multiple losses. To find weighting parameters for these losses, we cannot use cross-validation, because no anomaly examples are available to calculate anomaly detection quality. The work~\cite{perera2019ocgan} chooses weights empirically based on reconstruction quality. However, it requires a person to manually select the coefficients for each experiment, and it is not objective and reproducible.

\begin{algorithm}[t]
  \caption{Update gradient history}\label{algo:update_grad}
  \begin{algorithmic}[1]
    \Require \textit{net}, trained network, with calculated gradients of the loss function. \textit{history}, dictionary with previous values of gradient norm.
    
    \For{layer $\in$ net}
        \State history[layer.name] $\leftarrow$ history[layer.name] $\cup$ L2-norm(layer.weight.grad)
    \EndFor
\end{algorithmic}
\end{algorithm}

\begin{algorithm}[t]
  \caption{Select weighting parameter}\label{algo:select_weight}
  \begin{algorithmic}[1]
    \Require \textit{N}, the number of history points involved in the calculations. \mbox{\textit{history\_1}}, history information of 1'st loss. \mbox{\textit{history\_2}}, history information of 2'nd loss.
    \State weight\_per\_layer = new\_list()
    \For{layer $\in$ layers}
        \State values\_1 $\leftarrow$  select\_last\_N(N, exp\_smoothing(history\_1[layer]))
        \State values\_2 $\leftarrow$ select\_last\_N(N, exp\_smoothing(history\_2[layer]))
        \State weights\_per\_layer  $\leftarrow$ weight\_per\_layer  $\cup$ mean(values\_1 / values\_2)
    \EndFor
    \State weight $\leftarrow$ mean(weight\_per\_layer)
    \State \Return weight
\end{algorithmic}
\end{algorithm}

In order to choose weighting parameters automatically, we need to base our solution on measured values of an experiment.
Let $\bar{w} = [w_1, ..., w_n]$ be a vector of network parameters, $l_1(w)$ , $l_2(w)$ are losses calculated for this network, and 
\begin{equation}
L_{total}(w) = l_1(w) + \gamma l_2(w).
\end{equation}
Then 
\begin{equation}
\frac{\partial L_{total}(w)}{\partial w_i} = \frac{\partial l_1}{\partial w_i} + \gamma \frac{\partial l_2}{\partial w_i}.
\end{equation}
Depending on the nature of the loss functions $l_1$ and $l_2$, the norms of $\Vert \frac{\partial l1}{\partial w_i}\Vert$ and $\Vert\frac{\partial l_2}{\partial w_i}\Vert$ can differ by a factor of ten or even a hundred. Coefficient $\gamma$ regulates the relative influence of the loss functions in the total gradient with respect to this parameter $w_i$. To make the contribution of the loss functions equal, we can choose coefficient $\gamma$ in the following way: 
\begin{equation}
\Vert\frac{\partial l1}{\partial w_i}\Vert = \gamma \Vert\frac{\partial l_2}{\partial w_i}\Vert.
\end{equation}
However, due to using stochastic optimization, gradients are very noisy during training. To make this process more robust and stable, we propose to average the $\gamma$ coefficients over all network parameters and over their previous values (history information). Our approach is
summarized in Algorithm~\ref{algo:update_grad} and Algorithm~\ref{algo:select_weight}. 

In short: for each loss, we calculate the derivative (backpropagate loss) wrt each network weight $w_i$. Then for each convolutional layer we compute the L2-norm of the derivative wrt the weight matrix and store it. This is done after every $n_{weight}$ iterations in training, and all previously calculated values are kept, thus creating a gradient history per loss, per layer. We calculate the L2-norm per layer (but not per each weight $w_i$) to reduce the size of the stored information. Computing the norm over all network parameters would lose too much information, since the last layers usually have more parameters, 
and hence information about gradients from the first layers would be lost.
Firstly, the coefficient $\gamma$ is calculated per layer: we perform exponential smoothing of the history values of each loss (to make values robust to noise), and then calculate the average ratio between the last N entries in the gradient history for $loss_1$ and the same for $loss_2$. The final value for $\gamma$ is computed as the average over the $\gamma$-s per layer.

Our approach simply generalizes to a loss function consisting of more than two contributions. It also leaves room for research on which norm to use and how to compute the final weights.

%% file: sections/experiments.tex
\section{Experiments}

We show the effectiveness of the proposed approach by evaluation on six publicly available datasets and compare our method with a diverse collection of state-of-the-art methods for out-of-distribution detection, including state-of-the-art GAN-based approaches. 

\textbf{Datasets}. For evaluation we use the following well-known datasets (Table~1): \textbf{MNIST}~\cite{lecun2010mnist} and \textbf{Fashion MNIST (fMNIST)}~\cite{xiao2017fashion}, \textbf{COIL-100}~\cite{nene1996columbia} (images of 100 different objects against a black background, where views of each object are taken at pose intervals of 5 degrees), \textbf{CIFAR-10}~\cite{krizhevsky2009learning}, 
\textbf{LSUN}~\cite{xiao2010sun}
(we used only the \textit{bedrooms} and \textit{conference room} classes), and the \textit{aligned} \& \textit{cropped} face image attributes dataset \textbf{CelebA}~\cite{liu2015faceattributes}.

In all experiments, images of MNIST, Fashion MNIST and COIL-100 were resized to $32 \times 32$, examples of the LSUN dataset were downscaled to size $64 \times 64$ and for images of CelebA we made a $140 \times 140$ central crop and then downscaled to size $64 \times 64$. 

\begin{table}[b]
\setlength\tabcolsep{3.pt}
\renewcommand{\arraystretch}{1.2}
\centering
\begin{tabular}{c|c|c|c|c|c|c}
\hline
& MNIST & fMNIST & COIL-100 & CIFAR-10 & LSUN(bedr.) & CelebA \\
\hline
\# classes & 10 & 10 & 100& 10 & 1 & 40 attrib. \\
\# instances &  70,000 &  70,000 & 7,200  & 60,000 & 3,033,342 & 202,599\\
\hline
\end{tabular}
\vspace{2mm}
\caption{Statistics of the image benchmarks}
\label{table:dataset}
\end{table}

\textbf{Competing Methods}. As shallow baselines we consider standard methods such as \textit{OC-SVM}~\cite{chen2001one} and \textit{KDE}~\cite{parzen1962estimation}. We also test the performance of our approach against four state-of-the-art GAN-based methods:  \textit{AnoGAN}~\cite{schlegl2017unsupervised}, \textit{ADGAN}~\cite{deecke2018image}, \textit{OCGAN}~\cite{perera2019ocgan} and \textit{ALAD}~\cite{zenati2018adversarially}. 
Finally, we report the performance of three deep learning approaches from different paradigms: \textit{Deep SVDD}~\cite{ruff2018deep}, \textit{GPND}~\cite{pidhorskyi2018generative}, and the Latent Space Autoregression approach~\cite{abati2019latent} (results will
be reported under the name \textit{LSA}). All these methods have been briefly described in Section~\ref{sec:related}.

For ADGAN~\cite{deecke2018image}, OCGAN~\cite{perera2019ocgan}, ALAD~\cite{zenati2018adversarially}, GPND~\cite{pidhorskyi2018generative}, \textit{LSA}~\cite{abati2019latent} we used results as reported in the corresponding publications. Results for OC-SVM, KDE, AnoGAN were obtained from~\cite{abati2019latent}. 

\textbf{Evaluation Protocol}. To test the methods on classification datasets, we use a one-vs-all evaluation scheme, which has recently been increasingly used in anomaly detection papers~\cite{deecke2018image,perera2019ocgan,zenati2018adversarially,abati2019latent,ruff2018deep,pidhorskyi2018generative}: to simulate out-of-distribution condition, one class of a dataset is considered as normal data while images of other classes are considered as abnormal.
We evaluate results quantitatively using the area under the ROC curve (ROC AUC), which is a standard metric for this task.

\textbf{Implementation Details}. In all experiments we used pre-activation resnet blocks to build our generator, encoder, and discriminator; for computing relative-perceptual-L1 loss we used VGG-19~~\cite{simonyan2014very} network, pre-trained on Imagenet~\cite{russakovsky2015imagenet}. Other implementation details are presented in \textit{the supplementary material}.

\subsection{Results}

\input{sections/experiments/table_main.tex}

\textbf{MNIST and CIFAR-10}. Following~\cite{deecke2018image,perera2019ocgan,zenati2018adversarially,abati2019latent,ruff2018deep} we evaluated our approach on MNIST and CIFAR-10 using a train-test dataset split, where the training split contains part of the known class and the test split contains the unknown classes and the remainder of the known class.
We run each experiment 3 times with different initializations and present the averaged ROC AUC in Table~\ref{table:main}. 

Since the MNIST dataset is easy, all methods work well.
However, our approach allows to improve performance on several tricky digits (like 3 and 8) and outperforms all other approaches on average over the dataset. On the more diverse and complex CIFAR-10, the superiority of the proposed method is even more noticeable. This dataset contains 10 image classes with extremely high intra-class variation and the images are so small that even a human cannot always distinguish the kind of object on the image. Our approach based on the perceptual similarity of images can better capture class-specific information of an image, and hence, improves performance of anomaly detection in almost all experiments.

\textbf{fMNIST and COIL-100}. The work of \textit{GPND}~\cite{pidhorskyi2018generative} uses another train-test separation to evaluate performance. For a fair comparison, we repeat their evaluation scheme. The model trains on 80\% randomly sampled instances of a normal class. The remaining 20\% of normal data and the same number of randomly selected anomaly instances are used for testing. 
We report the average performance of GPND and PIAD on the fMNIST and COIL-100 datasets in Table~\ref{table:add_res} (left), along with OCGAN since they came out the second best on the previous datasets, and they report fMNIST/COIL-100 results in the same evaluation scheme as well. For the COIL-100 dataset we randomly selected one class to be used as normal, and repeated this procedure 30 times (as it was done in~\cite{pidhorskyi2018generative}). The comparison shows that PIAD excels on both datasets. 

\input{sections/experiments/table_add_res.tex}

\textbf{LSUN}. We also compare PIAD with the ADGAN approach on the LSUN dataset, training a model on images of bedrooms and treating images of the conference room class as anomaly. We achieve a ROC AUC of 0.781 against 0.641 with ADGAN. 

\textbf{CelebA}. In order to test our approach in conditions that are closer to a real-world use case, we experimented on the CelebA dataset, where we use attributes (Bald, Mustache, Bangs, Eyeglasses, Wearing\_Hat) to split the data into normal/anomaly cases. We train our model on 'normal' images where one of these attributes is \textit{not} present and test against anomaly images, where that same attribute \textit{is} present. Table~\ref{table:add_res} (center) shows the results. 

\begin{figure}[t]
    \centering
    \begin{subfigure}[]{0.33\textwidth}
        \centering
        \captionsetup{justification=centering, margin=0cm}
        \begin{subfigure}[]{0.95\textwidth}
            \centering
            \includegraphics[width=1\textwidth]{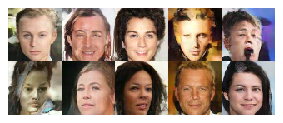}
        \end{subfigure}
        \caption{Random sample}
    \end{subfigure}%
    \begin{subfigure}[]{0.33\textwidth}
        \centering
        \captionsetup{justification=centering, margin=0cm}
        \begin{subfigure}[]{0.95\textwidth}
            \centering
            \includegraphics[width=1\textwidth]{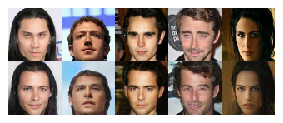}
            \caption{Reconst. of normal}
        \end{subfigure}
    \end{subfigure}%
    \begin{subfigure}[]{0.33\textwidth}
        \centering
        \captionsetup{justification=centering, margin=0cm}
        \begin{subfigure}[]{0.95\textwidth}
            \centering
            \includegraphics[width=1\textwidth]{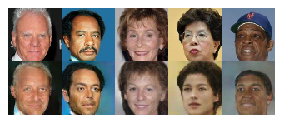}
        \end{subfigure}
        \caption{Reconst. of anomaly}
        \label{fig:exp_img_samples:an_rec}
    \end{subfigure}

    \caption{CelebA experiments. In (c) for Celeba each anomaly case is presented in sequence (from left to right):  Bald, Mustache, Bangs, Eyeglasses, Wearing\_Hat.}
    \label{fig:exp_img_samples}
\end{figure}

Anomaly attributes Eyeglasses and Wearing\_Hat are the easiest for PIAD. As shown in Figure~\ref{fig:exp_img_samples:an_rec} (4th and 5th examples), passing image $x$ through the encoder and the generator $G(E(x))$ removes glasses and hat from the images. Anomalies Mustache and Bangs are more of a challenge, but we  noticed that our model removes the mustache as well, and makes bangs more transparent. However, our model failed to recognize the Bald anomaly. This may be a result of the complexity of the anomaly (see Figure~\ref{fig:exp_img_samples:an_rec} first image, where indeed the man is not completely bald) and also inexact annotation (on Figure~\ref{fig:exp_img_samples:an_rec} the first image is annotated as bald, but the second is not).
 
\subsection{Ablation Study}

We also performed an ablation study on the CIFAR-10 dataset to show the effectiveness of each proposed component of PIAD. We considered 5 scenarios: \textbf{baseline}, i.e.\ our model with MSE as reconstruction error during training and as anomaly score, with empirically chosen weighting parameters by human supervision of the quality of generated examples and reconstructions; \textbf{+~gr-norm~w}, the same, but with gradient-normalizing weights policy; \textbf{+~perc}, where we further changed from MSE to perceptual loss; \textbf{+~perc-L1}, where we added normalization on $\mu, \sigma$ in perceptual loss and used L1-norm  over features instead of MSE, \textbf{+~rel-perc-L1},  where we used the proposed loss of Section~\ref{sec:perc}.  We present the average ROC AUC over CIFAR-10 in Table~\ref{table:add_res} (right).

\begin{figure}[tp]
    \centering
    \begin{subfigure}[]{0.33\textwidth}
        \centering
        \captionsetup{justification=centering, margin=0cm}
        \begin{subfigure}[]{0.95\textwidth}
            \centering
            \includegraphics[width=1\textwidth]{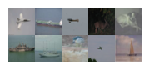}
        \end{subfigure}
        \begin{subfigure}[]{0.95\textwidth}
            \centering
            \includegraphics[width=1\textwidth]{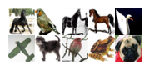}
        \end{subfigure}
    \caption{MSE: 0.481}
    \end{subfigure}%
    \begin{subfigure}[]{0.33\textwidth}
        \centering
        \captionsetup{justification=centering, margin=0cm}
        \begin{subfigure}[]{0.95\textwidth}
            \centering
            \includegraphics[width=1\textwidth]{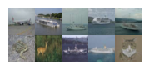}
        \end{subfigure}
        \begin{subfigure}[]{0.95\textwidth}
            \centering
            \includegraphics[width=1\textwidth]{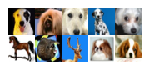}
        \end{subfigure}
    \caption{perc: 0.644}
    \end{subfigure}%
    \begin{subfigure}[]{0.33\textwidth}
        \centering
        \captionsetup{justification=centering, margin=0cm}
        \begin{subfigure}[]{0.95\textwidth}
            \centering
            \includegraphics[width=1\textwidth]{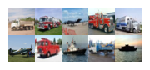}
        \end{subfigure}
        \begin{subfigure}[]{0.95\textwidth}
            \centering
            \includegraphics[width=1\textwidth]{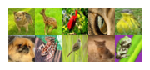}
        \end{subfigure}
    \caption{rel-perc-L1: 0.878}
    \end{subfigure}
    \caption{Least (top) and most (bottom) anomaly examples of car class of CIFAR-10. Results are presented with ROC AUC. 
    }
    \label{fig:cifar_ablation}
\end{figure}

We note that our proposed gradient-normalizing weight policy shows the same result as carefully found weights through parameter selection by a human after running the model several times. Each further modification improved results as well. Figure~\ref{fig:cifar_ablation} shows 
examples of images that were seen as the least and the most likely to be an anomaly, for different reconstruction losses.
We note that only relative-perceptual-L1 loss is not prone to select monochrome images as the least anomalous, and furthermore, this loss selected classes that are closest to the car classes as less anomalous: truck, ship.

%% file: sections/experiments/table_main.tex
\begin{table}[t]
\setlength\tabcolsep{2.5pt}
\begin{center}
\resizebox{0.8\textwidth}{!}{ 
\begin{tabular}{c c c cc c cccc c cc c c }
\toprule
&&\multicolumn{2}{c}{Shallow}&&\multicolumn{4}{c}{GAN-based methods}&&\multicolumn{2}{c}{Deep methods}&&\multirow{2}{*}{PIAD}\\
\cmidrule{3-4}\cmidrule{6-9}\cmidrule{11-12}
&& OC-SVM & KDE && AnoGAN & ADGAN & OCGAN & ALAD && LSA & Deep SVDD && (our)  \\ 
\cmidrule{1-14}
\parbox[t]{2mm}{\multirow{10}{*}{\rotatebox[origin=c]{90}{MNIST}}} &0&0.988&0.885&&0.926&\textbf{0.999}&\underline{0.998}&-&&0.993&0.980&&0.996\\
&1&\textbf{0.999}&0.996&&0.995&0.992&\textbf{0.999}&-&&\textbf{0.999}&0.997&&\textbf{0.999}\\
&2&0.902&0.710&&0.805&\underline{0.968}&0.942&-&&0.959&0.917&&\textbf{0.985}\\
&3&0.950&0.693&&0.818&0.953&0.963&-&&\underline{0.966}&0.919&&\textbf{0.981}\\
&4&0.955&0.844&&0.823&\underline{0.960}&\textbf{0.975}&-&&0.956&0.949&&\underline{0.960}\\
&5&0.968&0.776&&0.803&0.955&\textbf{0.980}&-&&0.964&0.885&&\underline{0.976}\\
&6&0.978&0.861&&0.890&0.980&0.991&-&&\underline{0.994}&0.983&&\textbf{0.995}\\
&7&0.965&0.884&&0.898&0.950&\underline{0.981}&-&&0.980&0.946&&\textbf{0.984}\\
&8&0.853&0.669&&0.817&\underline{0.959}&0.939&-&&0.953&0.939&&\textbf{0.982}\\
&9&0.955&0.825&&0.887&0.965&\underline{0.981}&-&&\underline{0.981}&0.965&&\textbf{0.989}\\
\cmidrule{3-14}
&avg&0.951&0.814&&0.866&0.968&\underline{0.975}&-&&\underline{0.975}&0.948&&\textbf{0.985}\\
\cmidrule{1-14}
\parbox[t]{2mm}{\multirow{10}{*}{\rotatebox[origin=c]{90}{CIFAR-10}}}
&airplane&0.630&0.658&&0.708&0.661&\underline{0.757}&-&&0.735&0.617&&\textbf{0.837}\\
&car&0.440&0.520&&0.458&0.435&0.531&-&&0.580&\underline{0.659}&&\textbf{0.876}\\
&bird&0.649&0.657&&0.664&0.636&0.640&-&&\underline{0.690}&0.508&&\textbf{0.753}\\
&cat&0.487&0.497&&0.510&0.488&\textbf{0.620}&-&&0.542&0.591&&\underline{0.602}\\
&deer&0.735&0.727&&0.722&\underline{0.794}&0.723&-&&0.761&0.609&&\textbf{0.808}\\
&dog&0.500&0.496&&0.505&0.640&0.620&-&&0.546&\underline{0.657}&&\textbf{0.713}\\
&frog&0.725&\underline{0.758}&&0.707&0.685&0.723&-&&0.751&0.677&&\textbf{0.839}\\
&horse&0.533&0.564&&0.471&0.559&0.575&-&&0.535&\underline{0.673}&&\textbf{0.842}\\
&ship&0.649&0.680&&0.713&0.798&\underline{0.820}&-&&0.717&0.759&&\textbf{0.867}\\
&truck&0.508&0.540&&0.458&0.643&0.554&-&&0.548&\underline{0.731}&&\textbf{0.849}\\
\cmidrule{3-14}
&avg&0.586&0.610&&0.592&0.634&\underline{0.657}&0.607&&0.641&0.648&&\textbf{0.799}\\
\bottomrule
\end{tabular}
}
\end{center}
\caption{ROC AUC for anomaly detection on MNIST and CIFAR-10. Each row presents an experiment in which this class was considered as normal data. For each line, the best result is shown in bold, and the second best result is underlined.}
\label{table:main}
\end{table}

%% file: sections/experiments/table_add_res.tex
\begin{table}[t]
	\begin{minipage}[t]{.31\linewidth}
	\centering
	\scalebox{0.9}{
    	\begin{tabular}[t]{cccc}
            \toprule
             & fMNIST & COIL-100 \\
            \cmidrule{1-4}
            GPND & 0.933 & 0.979 \\
            OCGAN & 0.924 & 0.995 \\
            \textbf{PIAD} & \textbf{0.949} & \textbf{1.000} \\
            \bottomrule
            \end{tabular}
	}
	\end{minipage}%
	~
	\begin{minipage}[t]{0.26\linewidth}
	\centering
	\scalebox{0.9}{
	    \begin{tabular}[t]{cc}
            \toprule
            & ROC AUC \\
            \cmidrule{1-2}
            Bald & 0.506 \\
            Mustache & 0.561 \\
            Bangs & 0.650 \\
            Eyeglasses & 0.777 \\
            Wearing\_Hat & 0.916 \\
            \bottomrule
            \end{tabular}
        }
	\end{minipage}
	~
	\begin{minipage}[t]{.4\linewidth}
	\centering
	\scalebox{0.9}{
	    \begin{tabular}[t]{cc}
            \toprule
            Model & ROC AUC \\
             \cmidrule{1-2}
            baseline & 0.609 \\
            + gr-norm w & 0.608  \\
            + gr-norm w + perc & 0.701 \\
            + gr-norm w + perc-L1 & 0.724 \\
            + gr-norm w + rel-perc-L1 & \textbf{0.799} \\
            \bottomrule
            \end{tabular}
	}
	\end{minipage}	
	\vspace{2mm}
	\caption{(left) Average ROC AUC for anomaly detection on fMNIST and COIL-100. (center) ROC  AUC  of  anomaly  detection  against  several  anomaly  classes on CelebA dataset. (right) Ablation study: average ROC AUC on CIFAR-10.}
	\label{table:add_res}
\end{table}

%% file: sections/conclusion.tex
\section{Conclusion}

We introduced a deep anomaly detection approach, built directly for the image domain and exploiting knowledge of perceptual image similarity. 
For the latter, we proposed a new metric that is based on perceptual loss, but is more robust to noise and changes of contrast of images.
As a part of our work we proposed an approach for selecting weights of a multi-objective loss function, which makes a contribution of all losses equal in the training process. We demonstrated the superiority of our approach  against state-of-the-art GAN-based methods and deep approaches of other paradigms on a diverse collection of image benchmarks. In the future, we plan to perform a more extensive evaluation of our method on higher resolution image data, like medical images.